\newcommand{\beq}{\begin{equation}}
\newcommand{\eeq}{\end{equation}}

\newcommand{\eb}{\begin{equation}}
\newcommand{\ee}{\end{equation}}

\documentclass[12pt,preprint,psfig]{aastex}

\usepackage{graphicx}
\usepackage{natbib}

\slugcomment{Accepted by the Astrophysical Journal, 21 Sept. 2011}

\shorttitle{Eta-sub-Earth Estimate from Kepler}
\shortauthors{Traub}

\begin{document}


\title{Terrestrial, Habitable-Zone Exoplanet Frequency from Kepler}


\author{Wesley A. Traub}
\affil{Jet Propulsion Laboratory, California Institute of Technology, Pasadena, CA 91109}
\email{wtraub@jpl.nasa.gov}





\begin{abstract} 
Data from Kepler's first 136 days of operation are analyzed to determine the distribution of exoplanets with respect to radius, period, and host-star spectral type.
The analysis is extrapolated to estimate the percentage of terrestrial, habitable-zone exoplanets.
The Kepler census is assumed to be complete for bright stars (magnitude $<14.0$) having transiting planets $>0.5$ Earth radius and periods $<42$ days. It is also assumed that the size distribution of planets is independent of orbital period, and that there are no hidden biases in the data. Six significant statistical results are found:
there is a paucity of small planet detections around faint target stars, probably an instrumental effect;
the frequency of mid-size planet detections is independent of whether the host star is bright or faint;
there are significantly fewer planets detected with periods $<3$ days, compared to longer periods, almost certainly an astrophysical effect;
the frequency of all planets in the population with periods $<42$ days is 29\%, broken down as terrestrials 9\%, ice giants 18\%, and gas giants 3\%;
the population has a planet frequency with respect to period which follows a power-law relation $dN/dP \sim P^{\beta - 1}$, with $\beta \simeq 0.71 \pm 0.08$; and
an extrapolation to longer periods gives the frequency of terrestrial planets in the habitable zones of FGK stars
as $\eta_\oplus \simeq (34 \pm 14)\%$.
Thus about one-third of FGK stars are predicted to have at least one terrestrial, habitable-zone planet.
\end{abstract}


\keywords{exoplanets, terrestrial, habitable zone, Kepler}




\section{Introduction}\label{intro}

The frequencies of exoplanet types, categorized by radius, period, and host-star spectral type, offer clues to the origin and evolution of exoplanet systems. Data from the initial 136 days of the Kepler mission are particularly valuable because they form a large and relatively complete sample, even at this early phase of operation. This paper examines the Kepler database to estimate the frequencies of each planet category, and extrapolates the data for an estimate of $\eta_\oplus$, the frequency of terrestrial planets in habitable zones around their host stars.

Since a major goal of the Kepler mission is to estimate $\eta_\oplus$, it is worthwhile early in the mission to analyze the data for this quantity. In addition, careful study of the data gives hints about how the mission is performing. All this information will be useful in guiding future decisions on data analysis methods and mission operations.

This paper discusses
the sample database (Sec.~\ref{data}),
bias estimation (Sec.~\ref{bias}),
the radius and period distributions in the sample (Secs.~\ref{radius}, \ref{periodobs}, \ref{pr}),
the radius and period distributions in the population (Secs.~\ref{period-under}, \ref{f-spty}, \ref{distnfns}),
and the habitable zone and estimated $\eta_\oplus$ (Secs.~\ref{HZ}, \ref{eta}).

\section{The Sample Database}\label{data}

In Borucki et al. (2011), hereafter ``B2011'', the database lists planetary candidates discovered during the first 136 days of observation by the Kepler mission.  In particular, Table 1 in B2011 lists the host star characteristics, and Table 2 lists the planetary candidates with their characteristics.  Hereafter the terms ``planets'' or ``exoplanets'' will be a shorthand for the more conservative term ``planetary candidates'' used by the Kepler team.

The combined database has 1235 planets.  For this paper, the following planets are removed: 16 labeled as single transits; 20 with host stars hotter than 6500 K; and 240 with hosts cooler than 5000 K.    For present purposes the stars are defined to have these temperature ranges ``K'' (5000-5499 K), ``G'' (5500-5999 K), and ``F'' (6000-6499 K). In each sub-sample the number of planets is F (159), G (475), and K (325), with a total FGK (959).  Using Table B1 from Grey (2008), these ranges correspond to the standard spectral types as follows: ``F'' $\approx$ F5-F9;  ``G'' $\approx$ G0-G7; and ``K'' $\approx$ G8-K2.  Hereafter, the quotation marks are dropped.

The number of target stars is estimated (http://archive.stsci.edu) using the following search qualifiers and values:
cadence (long cadence, 29.4 min.),
star radius ($<10$ solar radius), and
quarter (2nd).
The resulting number of target stars is 153,196, which agrees exactly with the number stated in B2011.
The FGK subset has 113,644 stars, about 74\% of the original sample.
The breakdown by spectral class is F (20,406), G (55,595), and K (37,643).
For perspective, the overall frequency of detection of all planets around FGK stars is then $959/153,196 \simeq 0.63\%$.

The star masses and surface gravities in the sample have ranges significantly larger than the narrow limits of textbook main-sequence dwarf stars, but are close enough to luminosity class V to be labeled as such.  Overall the 959-planet sample seems to be a good approximation of the target class often called ``Sun-like FGK stars''.

Hereafter $r$ refers to planet radius in units of $r_\oplus = 6378$ km, and listed in B2011 to the nearest 0.1.  The planet orbital period is $P$, in days, listed in B2011 to many significant figures.

Following standard statistical practice, the numbers of planets in the observed \emph{sample} are denoted by lower-case $n$, and the estimated numbers of planets in the parent \emph{population} are denoted by upper-case $N$.   Logarithms denoted by \textit{ln} are base e.

``Terrestrial, habitable-zone'' planets are defined here in terms of radius and surface temperature.  Terrestrial planets are taken to be those with $0.5 \leq r \leq 2.0$, corresponding to roughly 0.1-10 Earth masses (Lunine et al,, 2008).
For reference, the average radius of Uranus and Neptune is 3.9, Saturn is 9.4, and Jupiter is 11.18 (equatorial radius).
For convenience the dividing line between ice and gas giants is taken to be $r \leq 8$.
Thus three radius ranges are defined: small (terrestrial), medium (ice giant), and large (gas giant).

The habitable zone (HZ) is the region around a star where liquid water could exist on the surface of a planet.  This paper adopts three ranges of star-planet separation that have been proposed as HZ limits, as discussed in Sec.~\ref{HZ}.

References to bright or faint stars in this paper are a shorthand for apparent brightness or faintness, not absolute.

\section{Bias Estimation}\label{bias}

It is important to understand the biases that may exist in the database.  In this context, a bias is defined as a difference in character between the observed sample and the actual parent population.  As noted in Sec.~\ref{data}, the sample here is the set of transits and candidate planets in the Kepler database.  The parent population is the set of actual planets in orbits around Kepler's target stars.

In order to draw valid statistical conclusions, one must either explicitly compensate for a known bias, or make an assumption about the importance of a potential bias.  The bias analysis in this paper is based solely on the explanations in B2011, and on the initial assumption that all the data in the database are valid. No corrections are made for either over-counting  (e.g., false positives) or under-counting (e.g., missing events owing to poor detection in low signal-to-noise cases).  Also, no attempt is made to go outside the database itself, by using \textit{a priori} estimates of signal and noise, for example, in order to estimate the completeness of the period or radius data.

To be explicit, here is a list of some known or potential biases, and the corresponding assumptions to ignore them, or actions to mitigate them, as taken in this paper.  See the Kepler Data Release Notes (Christiansen, 2011) and Kepler Input Catalog (Brown et al., 2011) for explanatory details and extensive discussion of these and related points.

\textbf{Field-of-view bias.}  Are the Kepler stars and planets representative of the solar neighborhood, where we expect to find and characterize planets someday, or is there a bias owing to the different galactic location? The Kepler field of view (FOV) subtends a very small part ($~0.3\%$) of the sky, with a median target-star distance on the order of a kpc.  Therefore the Kepler target stars are certainly not in the immediate neighborhood of the Sun, so they may not be representative of the solar neighborhood.  However they are approximately at the Sun's distance from the galactic center.  Thus in terms of whether the Kepler population is representative of the solar neighborhood, there may be a bias, but it is usually assumed to be zero.

\textbf{Magnitude-limit bias.}  The Kepler sample is magnitude limited, not volume limited, so high-luminosity stars in the sample will tend to be farther away than low-luminosity ones.  However under the assumption that distant stars have the same statistical properties as nearby ones, per the FOV bias discussion above, and under the assumption that the Kepler stars are all at distances well within the galaxy, this too is assumed here to produce a zero bias.

\textbf{Active-star bias.}  Active stars have random brightness fluctuations on time scales that include transit times, adding noise to the photometric signal, reducing the likelihood of detecting a planet, especially a small one.  The Kepler team finds that giant stars have significantly more noise than dwarf stars (Christiansen, 2011), and that even dwarf stars have about $30\%$ greater photometric noise than expected on the basis of previous observations of the Sun (Dunham et al., 2011; Gilliland et al. 2011).  The result is that small planets are less likely to be found around active stars.  The effect of this noise is included in the transit detection algorithm because there is a signal-to-noise threshold requirement, but the bias against active stars is not compensated.  An \textit{a-posteriori} correction for this effect could be attempted after the Kepler mission is completed and more statistical information is in hand.  For the present paper, a zero bias is assumed.

\textbf{Star-spot bias.}  Noise from star spots is similar to active-star photometric noise, but at lower temporal frequency, so the active-star discussion applies here as well.  A zero bias is assumed in this paper.

\textbf{Stellar-parameter bias.}  The estimated planet radius depends directly on the assumed stellar radius, so any bias in the latter propagates to the former.  In the sense that a class of stars might tend to have a bias in stellar radius, the planets around those stars will be similarly biased.  Likewise the estimated planet semi-major axis depends directly on the assumed stellar mass, so similar considerations apply. Further, the assumed stellar luminosity affects the derived location of the habitable zone.  Limb darkening also will affect the derived planet radius, especially for near-grazing transits.  In this paper a zero bias is assumed for all stellar parameters.

\textbf{Spectral-class bias.}  A combination of other biases in this list against individual properties which define the spectral class of a star could result in an erroneous conclusion regarding, for example, the prevalence of planets around a given spectral type.  In the absence of evidence to the contrary, this paper assumes a zero bias.

\textbf{Impact-parameter bias.}  The probability of a transit depends on the assumed stellar radius and on the assumption that all transits across the disk are equally detectable.  However a grazing transit will generate a smaller photometric signature than an equatorial transit, so the effective stellar radius for calculating the probability of a transit will be less than the actual radius, and this ratio will depend on the signal and noise in a way that could be estimated.  For the present paper, the maximum impact parameter is taken to be the stellar radius, so a zero bias is assumed.

\textbf{False-positive bias.}  The elimination of false-positive detections from background eclipsing binary stars is a major consideration in the Kepler data pipeline.  This bias may tend to increase for fainter stars, owing to the low signal-to-noise in these cases.  It is possible that this type of bias exists in the present database, where an excess of giant planets is suspected around the fainter stars, as discussed in Sec.~\ref{radius-1}.  The present paper avoids this bias by considering only the brighter stars, and looking mainly for small-radius planets.

\textbf{Planet-radius bias.}  The existence of a finite noise level directly affects the detection threshold and therefore limits detection of the smallest planets.  It is likely that this effect exists in the present database, where a paucity of small planets is suspected around faint stars as compared to brighter ones (Sec.~\ref{radius-1}).  This is a type of bias that could be modeled in the future, when more is known about the actual noise in the data and the detection algorithm.  The present paper avoids this bias by basing its final results on the bright stars alone.  However there still could remain a bias against small planets around those stars as well, so in that sense the actual number of planets in the population might be larger than estimated here.

\textbf{Period-completeness bias.}  The Kepler team uses the rule that a minimum of three transits is required for a candidate planet detection.  If only one or two are seen, there is great uncertainty about the planet.  Likewise, if the detection algorithm does not perfectly adapt itself to the separate quarters of observations, between which the Kepler spacecraft rolls a quarter turn, and after which the stars fall on different detectors, then the likelihood of finding three consecutive transits is reduced.  This is the situation for the present database, although it probably will be remedied in future releases.  The present database does contain transits with periods greater than one-third of the relevant mission length.  However these transits were discovered in an \textit{ad hoc} manner, so there is no guarantee of completeness for these longer periods (B2011).  For this paper, only periods of less than 42 days are considered, although for completeness all periods are shown in some plots.  The restriction to periods less than 42 days is an important feature of this paper.

\textbf{Distribution-function bias.}  It is sometimes assumed that the frequency distribution of planets in the population can be modeled in terms of separable functions of spectral type, period, and radius (or mass).  This is a mathematical convenience that is allowable only because currently there is no strong theoretical or observational evidence to the contrary.  However as more data are accumulated from Kepler, radial velocity, and exoplanet microlensing observations, this convention will be tested and possibly replaced.  Nevertheless at present the bias introduced by this assumption is unavoidable, and it is so noted in this paper.

\textbf{Mission-length bias.}  The three-transit rule means that the Kepler mission length must be at least three times the length of the period of planets in the outer parts of the habitable zones, in order to fully sample those zones.  A shorter mission means that short-period data will need to be extrapolated to longer periods in order to estimate $\eta_\oplus$, for example.  Although this paper does carry out such an extrapolation (Sec.~\ref{eta}), there is uncertainty in doing so.  The bias incurred by extrapolation is entirely unknown, so in the present paper we merely note this uncertainty but do not attempt to make any corrections.

\section{Radius Distribution in the Sample}\label{radius}


\subsection{Radius bias: magnitude dependence}\label{radius-1}

\begin{figure}
\centerline{ \plotone{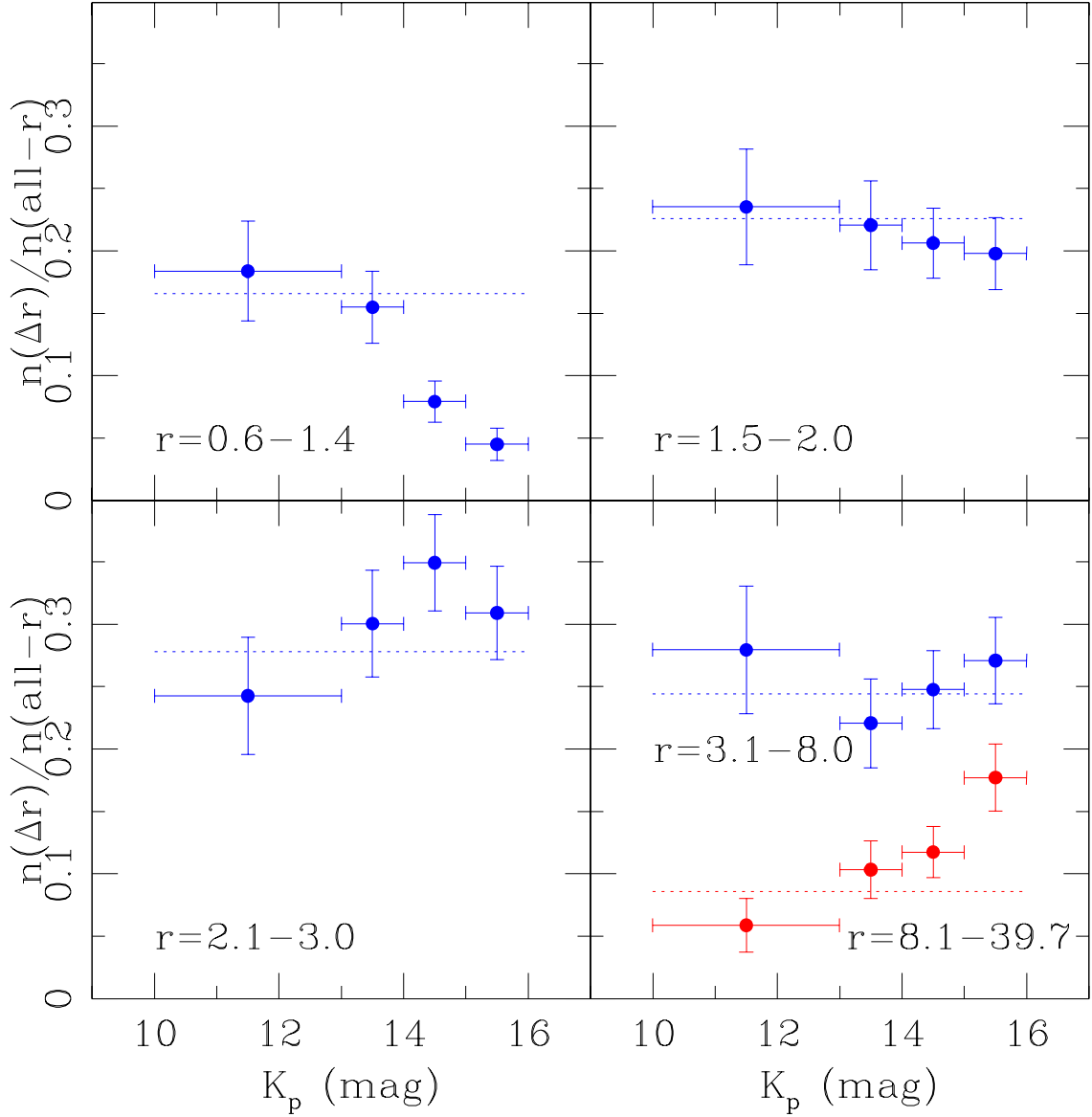} }
 \caption{ The fractions of stars in the sample with magnitude range from 10 to 16, and in the 5 radius groups (see inset values) from Table~\ref{table-nnkp}, are plotted along with Poisson error bars.  Each dotted line is an average from the bright ($K_p < 14$) stars in the radius group.  If there is a bias in the faint-star regime, it would be revealed by a disagreement between this line and the faint-star points. Thus, around faint stars, there appears to be a paucity of small planets, and an excess of large ones. }
 \label{fig-nnkp}
\end{figure}

As mentioned in B2011, there is a possible bias in the database owing to the fact that the signal to noise ratio decreases as the Kepler magnitude $K_p$ increases.  To search for a sign of this bias, the 959-planet sample is subdivided into 4 bins of target-star magnitude, with 136 stars in the $K_p$ range $(10, 12.999)$, 213 stars in the range $(13.0, 13.999)$, 315 in the range $(14.0, 14.999)$, and 288 in the range $(15.0, 15.999)$.  The 6 stars brighter than 10.0 and the 1 star fainter than 16 are ignored.  In each magnitude range, planets in 5 radius ranges are counted.  The bins in magnitude and radius are chosen to give roughly similar numbers in each category, to aid statistical comparison.  The data are listed in Table~\ref{table-nnkp}, and the fraction $n(\Delta r)/n(\textrm{all-r})$ in each radius group $\Delta r$ is plotted in Fig.~\ref{fig-nnkp}.

In the plots for the middle three groups (radii 1.5-2.0, 2.1-3.0, and 3.1-8.0), the fraction of planets in the sample is approximately constant in going from Kepler magnitude 10 to 16, as judged by the overlap or near-overlap of the error bars in each sub-group.  However for the smallest and largest planets the case is different.

For the smallest planets (radii 0.6-1.4), there is a highly significant drop in planets detected around the faintest stars ($K_p = 14$ to 16) compared to the numbers found around brighter stars.  This is a clear sign of the radius bias mentioned in B2011.  Quantitatively, from this figure it appears that the break point is close to $K_p = 14$.  For convenience in this paper, ``bright'' Kepler stars are defined as those with $K_p < 14.0$, and ``faint'' stars are defined as those with $K_p \geq 14.0$.  The bright star sample may still be incomplete in terms of the smallest planets, but in this paper the sample is assumed to be complete.  The faint sample is not complete, and therefore will be ignored for the purpose of estimating numbers of planets in the population (Secs.~\ref{period-under}-\ref{eta}).

In each panel of Fig.~\ref{fig-nnkp} the average number of stars in the bright group is indicated by a horizontal dotted line.  If there is no bias, then the faint groups should lie within about $1 \sigma$ of the dotted line. The middle three radius groups are seen to be consistent with this average (see also Sec.~\ref{radius-3}), but the smallest radius planets around faint stars are seen to be 5 to 10 $\sigma$ below that line, and are therefore highly significant.

For the largest planets (radii 8.1-39.7), there are significantly more planets detected around faint stars than bright ones.  There is no obvious astrophysical reason for this effect, although one explanation might be that false positives are being picked up from background eclipsing binaries.  It could be more difficult to differentiate eclipsing binaries from planetary transits when the target star is relatively faint. If this apparent excess is indeed the case, then the detection of about 41 events out of 959 suggests that the rate of unrecognized false positives is around $4\%$;  this is much lower than the value of about $50\%$ mentioned in the original data release (Borucki et al., 2010), but more in line with what B2011 says is the ``substantially smaller'' rate expected in the current database.

\subsection{Radius bias: mid-size planets}\label{radius-2}

This section extends the analysis in Sec.~\ref{radius-1} of the absolute numbers of mid-size planet transits as a function of star magnitude,
to ask if the relative numbers in the sample have any dependence on host star brightness.  Column 8 in Table~\ref{table-nnkp} shows the basis number of stars $N_{\star}$ observed by Kepler in each magnitude range from 10 to 16.  Periods are limited to 42 days.
The mid-size planets, those with radii in the range 1.5 to 8.0, should be free of the apparent bias at the small and large ends of the radius scale. The ratio of the number of these well-measured planets to the basis number of stars is listed in the last column of Table~\ref{table-nnkp}, along with the Poisson uncertainty.

Three of the 4 magnitude groups have excellent agreement on the number of detected planets per star, consistent with an average of $(0.44 \pm 0.04)\%$, within the uncertainties.  In these 3 groups there does not appear to be any trend, and certainly not a significant trend toward fewer detections at faint magnitudes, as one might expect.  However in the remaining group, for magnitude-14 stars, the number of planets jumps up to $(0.64 \pm 0.04)\%$, well above the average of the other groups.  Averaging the two faint bins together gives a ratio of $(0.50 \pm 0.02)\%$, which is just within $1\sigma$ of the bright group average of $(0.46 \pm 0.03)\%$, so it appears that there is no evidence for a bias against detection of the mid-range of planet radii, 1.5 to 8.0 Earths, when comparing bright and faint target stars.  The overall frequency of mid-size planet detection is $(0.49 \pm 0.02)\%$ for all FGK bright and dark targets combined.  For comparison, this is smaller than the frequency of detection of all planet sizes in the sample, $0.63\%$, from Sec.~\ref{data}, the difference being that the smallest and largest planets are not included.

\subsection{Radius bias: bright \textit{vs} faint stars}\label{radius-3}

Since the analysis in Sec.~\ref{radius-1} showed that there is a fairly well-defined transition at $K_p \simeq 14.0$ beyond which small planets appear to be incompletely sampled, it is worthwhile to look at the overall radius distribution in the sample and to see how it depends on the bright and faint regimes.

\begin{figure}
\centerline{ \plotone{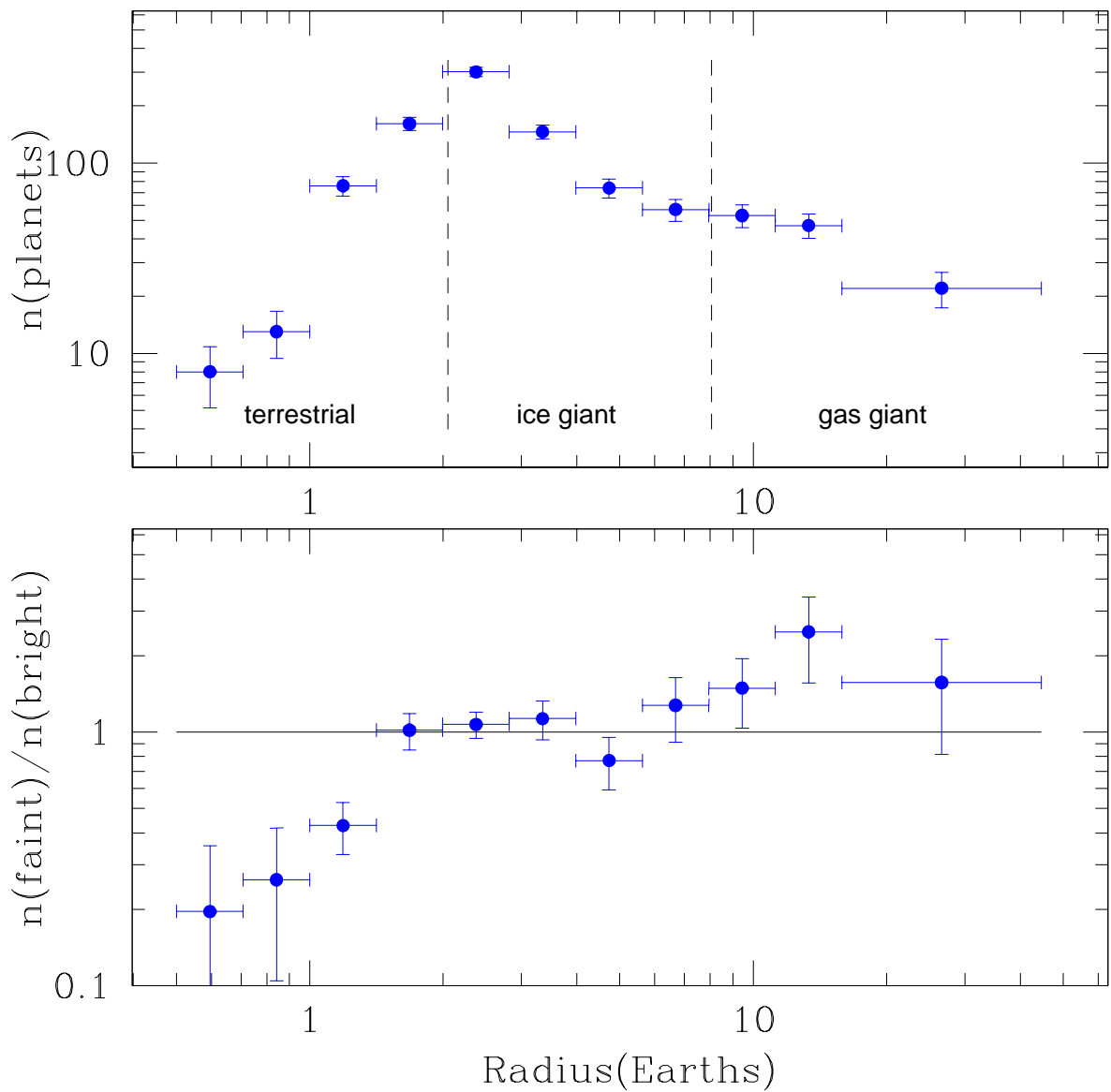} }
 \caption{ (upper) The numbers of planets in the sample are shown as a function of radius, with Poisson uncertainties.
 For reference, the nominal planet type is indicated for each radius range: terrestrial, ice giant, and gas giant.
 (lower) The ratios of numbers per bin for faint/bright host stars are shown, normalized to the totals of each.  The paucity of small planets in the faint group is seen as a strong drop in this ratio in the 3 smallest-radius bins.  The slight excess of large planets around faint stars, in the 3 largest-radius bins, is a possible indication of unrecognized false positive detections.  }
 \label{fig-nr}
\end{figure}

To do this, the 959 planets in the FGK database, with periods less than 42 days, are binned into bins of equal size in $log(r)$ space, in steps of $\Delta log(r) = 0.15$, anchored at $r = 1$, and listed in column 2 of Table~\ref{table-r}.  The breakdown into 355 bright and 604 entries is shown in columns 3 and 4.  The radius data are visualized in Fig.~\ref{fig-nr}, where the upper panel shows the total number (bright plus faint) of planets in each radius bin.

To see if there is any bias in the sample, in going from bright to faint targets, the lower panel in Fig.~\ref{fig-nr} shows the ratio faint/bright in each radius bin, normalized to the total number in each range, along with Poisson error bars.  For reference, note that across the mid-radius ice-giant group, and for one bin on either side, the ratio is essentially flat within the noise; this shows that Kepler is detecting planets around bright and faint stars equally well, across this range of planet radii.  This is in agreement with Sec.~\ref{radius-2}.  However there are exceptions at the small- and large-planet ends of the distribution, as discussed next.

As was seen in Sec.~\ref{radius-1}, many more small planets ($r < 10^{0.15} = 1.41$) are detected around (apparently) bright stars than faint ones.  There is no astrophysical reason for this to happen, unless somehow there is a spectral-type bias in the detections, which cannot be discerned from the current data alone.  The most likely reason for this difference is that small planets around faint stars are being missed by the data analysis algorithm.  In this range, a total of 38 small planets around faint stars are detected, whereas about $100 \pm 13$ should have been seen, based on the bright-star numbers.  This suggests that the detection efficiency for small planets around faint stars is only about $(38 \pm 6)\%$ of the efficiency around bright stars.

To conclude this section, the data show that the database is biased against small planets ($r < 1.4$) around faint stars
($K_p \geq 14.0$), so for the remainder of this paper the sample basis will be the bright star ($K_p < 14.0$) subset.
There are 35,896 FGK target stars in this bright sample, subdivided by spectral type as F (11,819), G (14,997), and K (9080); these are the basis numbers of target stars in the bright-star population.

\section{Period Distribution in the Sample}\label{periodobs}

\begin{figure}
\centerline{ \plotone{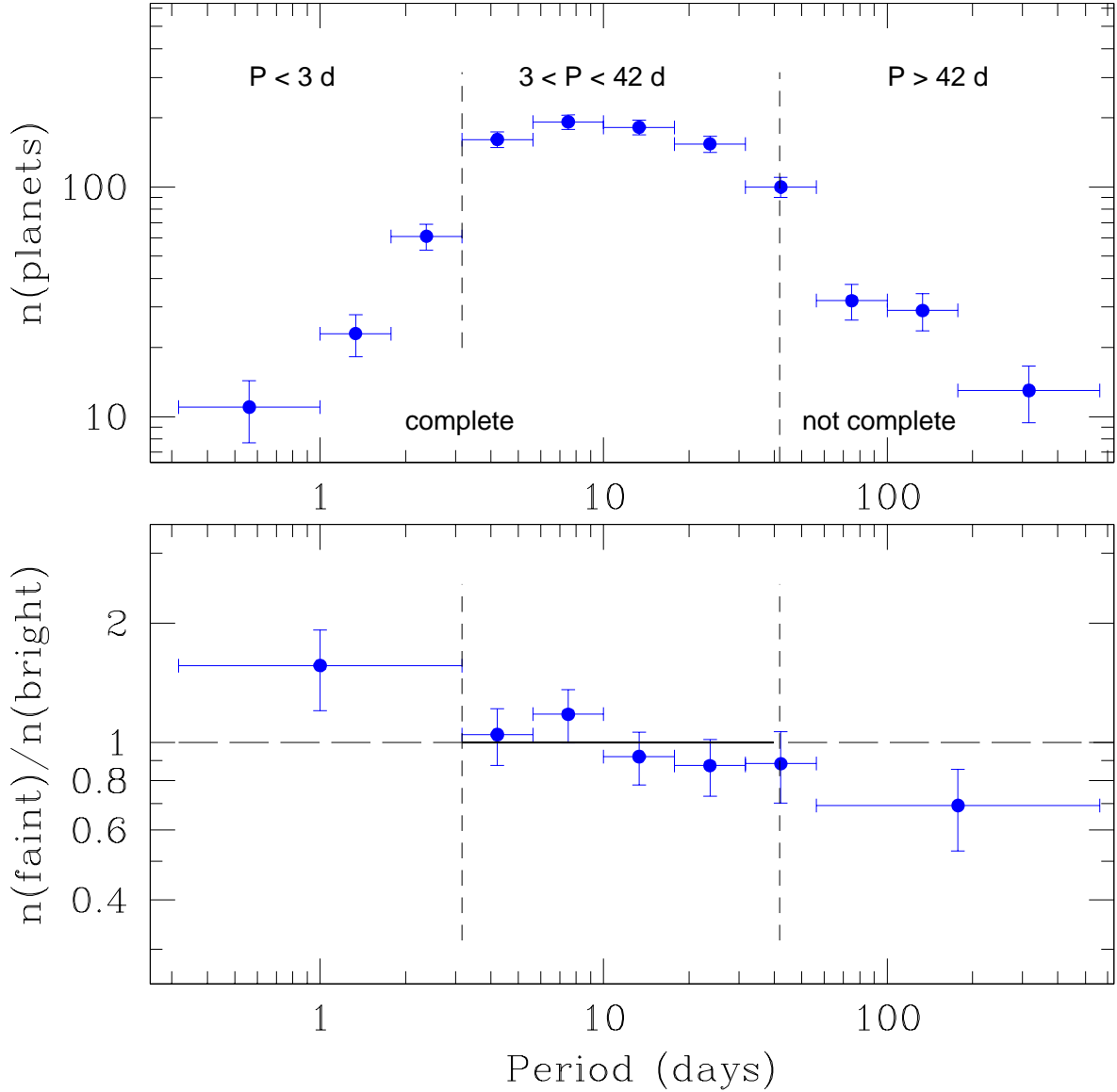} }
 \caption{ (upper) The numbers of planets detected in each period bin in the sample are shown, with Poisson uncertainties.
 For reference, the nominal period ranges are indicated: for $P<3$ days, the sample is complete, so the apparent drop-off is astrophysical in origin; for $3<P<42$ days, the sample is also complete; for $P>42$ days, the sample is not complete, and may be biased, so the drop-off is likely an artifact.
 (lower) The ratios of numbers for faint/bright host stars are shown, normalized to an average of unity.
 Within the completely-sampled range ($P<42$ days), there does not appear to be any bias from faint targets compared to bright ones.  However the apparently systematic trend toward a relatively smaller number of long-period planets around faint targets, compared to bright ones, is a possible bias at the $1 \sigma$ level. }
 \label{fig-np}
\end{figure}

The numbers of planets in each interval of $\log(P)$, where $P$ is the planet orbital period in days, are listed in Table~\ref{table-p}, in bins of size $\Delta \log(P) = 0.25$.  As in the radius discussion, the numbers for all Kepler magnitudes are listed along with the breakdown into bright and faint stars.  The total number is 958, one less than for the radius listing, because one very long-period planet is dropped.  The same data are plotted in the upper part of Fig.~\ref{fig-np}, where 3 period regimes are indicated.

For short periods, $P < 3$ days, there is a sharp drop-off, which almost certainly is an astrophysical effect, since these planets would have had many transits in the database and would be relatively easy to detect. There is a mild potential bias against short period detections in the sense that individual transits get shorter as the period decreases, however this is compensated by the fact that there are more of them to count; the net effect varies slowly with period (cf. Sec.~\ref{pr}), certainly much slower than the abrupt drop-off seen in Fig.~\ref{fig-np}.

For periods in the range from 3 to 42 days, the current database is expected to be statistically complete, since at least 3 transits (a Kepler requirement) should have been detected in the database's 136-day window.

For longer periods the efficiency of detection in the current database is expected to drop, because B2011 notes that periods greater than 42 days were not searched for in a systematic fashion.  Therefore the fall-off for long periods should be no surprise, since there is certainly a selection effect here, with no implied astrophysical meaning.

The normalized ratio of detections in the sample, $n(\textrm{faint})/n(\textrm{bright})$, is plotted in the lower part of Fig.~\ref{fig-np}, similar to the plot for the radius distribution.  For short periods, the numbers are small, so the error bar is large, and there is no obvious interpretation.  In the range where the data are complete and abundant, 3-42 days, the faint and bright data sets are identical in relative numbers of detections, within the counting statistics.  Indeed, there is no obvious reason why there should be any kind of instrumental or astrophysical bias here.  The slow downward drift of the ratio, as the period increases, is slightly puzzling; this may indicate a difficulty in detecting long-period transits in the fainter stars, which would not be a surprising instrumental bias, but the significance is low, and more data will be needed to see how real this is.

\section{Period-Radius Scatter Diagram}\label{pr}

\begin{figure}
\centerline{ \plotone{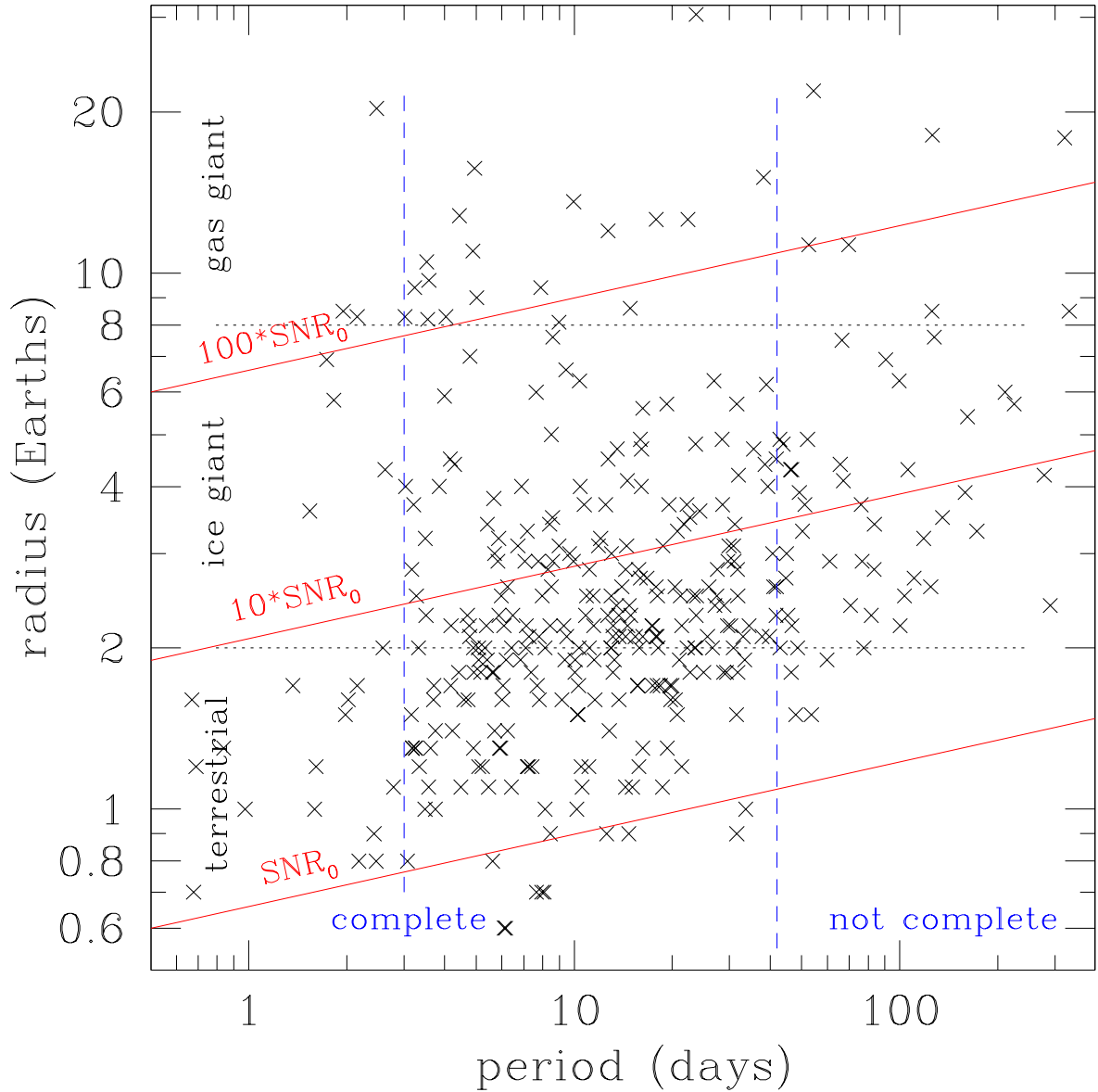} }
 \caption{ The period and radius of Kepler planets in the sample, around bright stars, are plotted.
 The lower right corner is relatively empty, probably owing to low SNR there, not because small planets are absent from long periods.
 The upper left corner is relatively sparse, in spite of an expected high SNR there, implying a deficit of large planets on short-period orbits.
 The left side of the diagram is relatively empty owing to an apparent paucity of planets of all sizes at periods less than 3 days.
 The right side of the diagram is not completely sampled in the current database, so should be ignored here.     }
 \label{fig-pr}
\end{figure}

The possibility of a correlation between period and radius can be investigated by plotting the 355 Kepler planets around bright stars in a (period, radius) scatter diagram, as shown in Fig.~\ref{fig-pr}.  To guide the eye, a vertical line at $P = 3$ days isolates the short-period range where planets are apparently not frequent (Sec.~\ref{periodobs}).  Another vertical line at $P = 42$ days indicates the cutoff point, above which the database is not complete, and which is now ignored.  A pair of horizontal lines at $r = 2$ and 8 Earth radii arbitrarily divides the diagram vertically into small-, medium-, and large-radius planets.

The diagonal lines are a crude guide to the regions of relatively easy vs relatively hard planet detection as a function of $(P,r)$.  The simple assumption here is that the number of planets will depend on the signal to noise ratio (SNR) of the transits.   The signal for a single transit is proportional to the transit time and the ratio of planet to star area, hence $P^{1/3}r^2$.  The noise is proportional to the square root of the transit time, hence $P^{-1/6}$.  The SNR for multiple transits is proportional to the square root of the number of transits, or $P^{-1/2}$. The net SNR for a given mission length is $\textrm{SNR} \sim {{\textrm{SNR}}_0} \cdot P^{-1/3} r^2$, where ${\textrm{SNR}}_0$ is a factor that depends on the star flux, etc., but not $P$ or $r$.  Thus lines of constant SNR can be drawn as $r = \sqrt{\textrm{SNR/SNR}_0} \cdot P^{1/6}$.  These lines are drawn for the cases of 1, 10, and 100 times ${\textrm{SNR}}_0$, where ${\textrm{SNR}}_0$ is arbitrary.  Several features of the scatter diagram are immediately explained by this simple SNR argument, as follows.

First, the area below the ${\textrm{SNR}}_0$ line, to the lower right, appears to be relatively empty of planets, and this is to be expected; it should not be concluded that there are fewer planets with small radius at large period, for example, because this region is simply the one where detection is the most difficult.

Second, the area toward the upper left is one where detection should be very easy, with many transits of large-radius planets, however the region is relatively empty.  This indicates that there truly are very few planets in this region, i.e., large planets on short periods are rare. This is the opposite of the early indications from radial velocity where there appeared to be a pileup of large planets on short period orbits, which was seen even then as a possible bias of that technique, and is shown clearly here.

Third, within the central vertical strip, between the 3- and 42-day lines, the relative density of points appears to be approximately uniform in $\log(P)$, however to the right of the 42-day line the density of points drops off rapidly.  This simply illustrates that the sampling is not necessarily complete for this long-period region; there is not necessarily any astrophysical meaning to this drop-off.  This region will be better sampled as the Kepler mission progresses in time, and more completely-sampled period data is released.

Fourth, the trend of data points in the center 3-42 day region appears to be slightly upward in slope, roughly parallel to the SNR lines.  In fact, a fit of the median radius in 6 equally-spaced bins of $\log{P}$ (not shown) reveals that the apparent median radius varies as $P^\gamma$ where $\gamma \simeq 0.11 \pm 0.05$, which is consistent at about the $1\sigma$ level with a slope of $1/6 \simeq 0.17$.  The similarity of slopes suggests that the trend is purely an artifact of the detection method, and not likely to be of astrophysical relevance.

\section{Period Distribution in the Population}\label{period-under}

For every transiting planet, there are many more non-transiting planets.  It is well known that the probability of a transit is simply $p_t = R(star)/a(orbit)$.  Before the Kepler mission was launched, a massive effort was invested in characterizing the target stars (Brown et al., 2011), one benefit of which is that the Kepler database now contains \textit{a priori} estimates of the host star mass and radius, and of course semi-major axis $a(orbit)$ from the period and star mass.

In a statistical sense, for every transit there are a total of $1/p_t$ planets in transiting plus non-transiting orbits.  Thus it is easy to estimate $N_p$, the total number of planets in the population orbiting the observed stars, simply by counting the observed planets $n_p$ with a weight factor of $1/p_t$, giving
\begin{equation}\label{pt}
  N_p = \Sigma_{i=1}^{n(obs)} (p_t(i))^{-1} .
\end{equation}
This value, the number of planets per bin in the population, is listed in Table~\ref{table-p-under} as a function of period, for the planets around bright FGK target stars.

\section{Frequency and Radius vs Spectral Type in the Population}\label{f-spty}  

The original question, ``what is the frequency of planets in the target population?'', can now be addressed.  To minimize the effect of biases in the data sample, only bright Kepler stars are considered, and of those, only ones with planet periods less than 42 days.  There are sufficient data to break down the planets by radius into the terrestrial, ice giant, and gas giant groups discussed above.  And the spectral types of stars are broken down into F, G, and K groups, also discussed above.  For each of these nine sub-groups, the number of planets in the population can be estimated by assigning a projection factor $1/p_t$ to each observed planet in the sample, and summing over the projected estimates, using Eqn.~\ref{pt}.  The resulting numbers of planets in the sample, $n_p$, and in the population, $N_p$, are tabulated in Table~\ref{table-pssp}
The total number of stars with transits $n_s$ in the sample, and the number of stars in the population $N_s$ are also listed.  The bottom row in the table gives the sums of entries above in each case.

As a simple check, note that the $N_p$ entries are roughly a factor of 100 larger than the $n_p$ entries, which is appropriate because the average transit probability is roughly $\bar{p_t} \sim 1/100$.  However the numerical value depends on the exact orbit and star size, so the factor varies from one system to another.  As a direct result of this variation, it should be expected that some global quantities will be different depending on whether it is the sample or the population that is being considered.  As an example, the relative number of terrestrial planets in the sample is $140/355 \simeq 0.394$, whereas this ratio in the population is $3073/10571 \simeq 0.291$, which is significantly smaller; the latter value will be needed in Sec.~\ref{eta}.

The same data is displayed as percentage ratios of planets to stars, e.g., $N_p(terr)/N_s$, etc., in Table~\ref{table-f-spty}.  The error bars in this table are derived from the Poisson statistics of the $n_p$ values, i.e., $(N_p/N_s)\cdot \sqrt{n_p}/n_p$.  The actual errors will be larger, owing to the fluctuations expected from the projection process as applied to a small sample, as discussed above.  Nevertheless, it is of interest to draw some tentative conclusions from Table~\ref{table-f-spty}, although these may change as more Kepler data become available.

One conclusion is that the fraction of stars with terrestrial-radius planets (and in short $P < 42$ days) is approximately the same for F, G, and K stars, at about 9\%.  On the other hand the fraction of ice giant planets varies by nearly a factor of two, being about 14\% for F an K stars, but 24\% for G stars; if this trend holds for longer-period planets, it may be a clue about planetary origin and evolution.  Finally, for the gas giants, the fraction of stars with these planets, again in $P<42$ day orbits, is a rapidly-dropping function of spectral type, going from 5\% around F stars to 2\% around K stars; since it is conceivable that giant planets may tend not to form around lower-mass stars, this too will be of interest to follow as more data become available.

The last column of Table~\ref{table-f-spty} shows that the number of all planets (in short orbits) per star is roughly constant at about 29\%, independent of spectral type, so short-period planets are a relatively common phenomenon.

\section{Period Distribution Model}\label{distnfns}

It is useful to have a parameterized model of the frequency of occurrence of planets, as functions, for example, of host star spectral type, and planetary mass, radius, and period.  A model can facilitate comparison with theories of the evolution of planetary systems, and also, as in this paper, for estimating the frequency of planets beyond the current range of measurements.

For the present data set, the lack of correlation between radius and period in Fig.~\ref{fig-pr} suggests that the frequency distribution in terms of radius is independent of the frequency distribution in terms of period.  Also, the approximately constant value of planet frequency with respect to spectral type (Sec.~\ref{f-spty}) suggests a possible lack of correlation here too.  Thus a model in which the distribution function is represented by a product of functions of radius and period, respectively, seems appropriate.

The essentially monotonic increase in the estimated number of planets in the population, with increasing logarithmic period, in Table~\ref{table-p-under}, suggests that a power law in period could be an appropriate model.  Using $f(P)$ to denote the ratio of planets to stars, or essentially the average number of planets per star, a power law of the form
\begin{equation}\label{power-law}
  \frac{df}{d\ln P} = A P^\beta
\end{equation}
or equivalently
\begin{equation}
  \frac{df}{dP} = A P^{\beta - 1}
\end{equation}
seems appropriate.

To fit this trial law to the data at hand, the data first need to be cast into an appropriate form, as follows.
The data points to be fit are those from a discrete version of $df/d\ln P$, written here as $\Delta f / \Delta \ln P$.
Using data from Table~\ref{table-p-under}, the discrete number of planets in the population in the \textit{i-th} bin, $\Delta N_p(i)$, divided by the number of target stars, $N_s$, can be written as
\begin{equation}
  \Delta f(i) = \frac{\Delta N_p(i)}{N_s}.
\end{equation}
The basis number of bright target stars is $N_s = 35,896$, from Sec.~\ref{radius-3}.
The $\Delta \ln P$ term can be written as
\begin{equation}
  \Delta\ln P(i) = \ln P_{i+1} - \ln P_i = 0.4609
\end{equation}
which in the present case is constant for all intervals.  Thus the data to be fitted to the model are the values of $\Delta f / \Delta \ln P$ in each period bin; these values are listed in Table~\ref{table-p-under}.

The data are fitted by taking the logarithm of both sides of Eqn.~\ref{power-law}, to cast the model in the form of a linear equation $y = a + bx$, and a weighted least-squares algorithm used to obtain the coefficients, where the weights are obtained from the uncertainties in the number of planets per bin ($n$) in the sample, so $\Delta N = N \Delta n / n = N/\sqrt{n}$.  The 6 fitted bins are from rows 4 through 9 in Table~\ref{table-p-under}, i.e., those with periods greater than about 3 days, given that there is an apparent fall-off in numbers below this point (Sec.~\ref{periodobs}), and those with periods less than about 42 days, given that the database is not complete above this point (Sec.~\ref{periodobs}).  The fitted parameters are
\begin{equation}\label{abeta}
    A = 10^{-1.99 \pm 0.09} \simeq 0.0103 \pm 0.0022
\end{equation}
and
\begin{equation}\label{beta}
    \beta = 0.71 \pm 0.08 .
\end{equation}
The reduced chi-square value is ${\chi_{red}}^2 = 10.75/(6-2) = 2.7$ which suggests that the data have more uncertainty than given by Poisson statistics and/or the model is not optimum; at this point, only more data will help resolve these issues.

The $df/d\ln P$ data and model results are plotted in Fig.~\ref{fig-npunder3}, where a thick line indicates the model over the fitted period range, and extensions of the model to shorter and longer periods are shown as thinner dashed lines.  Horizontal error bars indicate the widths of the individual bins, and vertical error bars indicate the Poisson uncertainties, mentioned above.  It is clear that there is nominal agreement between the data and model, and that the degree of complexity of the model (2 parameters) as well as its functional form appear to be apropriate for the data at hand.  Future data will be absolutely crucial in determining the robustness of the present model.

\section{Habitable Zone}\label{HZ}

There is general agreement that the HZ is defined as the planet-star distance range within which liquid water can exist on a planet's surface.  The surface temperature of a planet is a function of stellar luminosity, albedo, greenhouse effect, eccentricity, obliquity, rotation rate, and geologic age.   Of these, only the first parameter can be estimated for the Kepler planets.  To encompass the effect of the remaining parameters, this paper adopts three ranges that have been proposed to date, summarized in the first three columns of Table~\ref{table-HZ}, and all specified for the case of the Sun.

Case 1, a ``wide'' HZ, 0.72 to 2.00 AU, covers a generous range of semi-major axis values, from Venus (0.72 AU) to Mars (1.52 AU) and beyond, since Venus may have had liquid water at one time, before it entered a runaway greenhouse phase, and because Mars almost certainly had liquid water at one time.  With a more effective greenhouse, a planet even farther from the Sun, out to 2.0 AU, also may have had liquid water.  This is the range recommended to the Kepler team in order ``to be sure not to exclude planets that could conceivably be habitable'' (J. Kasting, priv. comm., 2011).

Case 2, a ``nominal'' HZ, 0.80 to 1.80 AU, is somewhat more restrictive, with an inner edge between Venus and Earth, but with the outer edge still slightly beyond Mars, reflecting less extreme assumptions than the first case.  This is the range that was recommended for the TPF-C project (Levine, Shaklan, and Kasting, 2006), and is ``a `best bet' estimate for the HZ'' (J. Kasting, priv. comm., 2011).


Case 3, a ``narrow'' HZ, 0.95 to 1.67 AU, tightens up even more on the previous cases, reflecting a more conservative view.  This is the range that will give ``a lower limit on $\eta_\oplus$, so that you're sure to build your TPF telescope big enough'' (J. Kasting, priv. comm., 2011).

The corresponding orbital periods are estimated as follows.  For circular orbits the HZ distances $a_\odot(in)$ and $a_\odot(out)$ are scaled with stellar luminosity $L$  as $a \sim L^{0.5}$. For non-solar stars the luminosity is modeled to vary as $L \sim M^{3.8}$, where $M$ is stellar mass.  From Kepler's law, $P^2 \sim a^3/M$, which, after substituting, gives
\begin{equation}
  P = 365.25 \  M^{2.35} a_\odot^{1.5} .
\end{equation}
Here $P$ is in days, $M$ is in solar masses, and $a_\odot(AU)$ is the inner or outer edge of the HZ for the three cases listed.
The median star masses in the Kepler database are 1.13 (F), 1.08 (G), and 1.01 (K).  The resulting period ranges for each case and spectral type are listed in Table~\ref{table-HZ}.

\section{Eta-sub-Earth}\label{eta}

The average number of planets per star $(f_2 - f_1)$ in a period interval $(P_1, P_2)$, in the power-law model of Eqn.~\ref{power-law}, is obtained by integration, giving
\begin{equation}
  f_2 - f_1 = \frac{A}{\beta}(P_2^\beta - P_1^\beta) .
\end{equation}
To specialize to terrestrial planets, this should be multiplied by $\rho_\oplus$, the ratio of terrestrial planets ($r = (0.5, 2.0)$) to all planets, where
\begin{equation}
  \rho_\oplus = \frac{N_p(terr)}{N_p}  \simeq 0.291
\end{equation}
for short-period planets around bright stars, from Sec.~\ref{f-spty} and Table~\ref{table-pssp}.
Thus the average number of terrestrial planets per star, in the population, as a function of spectral type and HZ range,
is $\eta_\oplus(\textrm{SpTy, HZ})$, where
\begin{equation}
  \eta_\oplus(\textrm{SpTy, HZ}) = \rho_\oplus \cdot (f_2 - f_1) = \rho_\oplus \cdot \frac{A}{\beta} \cdot (P_2^\beta - P_1^\beta)
\end{equation}
The estimated values are given in Table~\ref{table-eta-terr}, where the range is from a low of 22\% to a high of 47\%.
To be clear, these estimates are based on projecting the total of all planets around all bright stars in the database, then simply applying the terrestrial fraction for short periods to the longer HZ periods; the individual spectral classes were not fitted, only the sum was fitted.

To obtain a single value for the number of planets per star, averaged over spectral class and HZ size, the entries in Table~\ref{table-eta-terr} are averaged to give
\begin{equation}
  \eta_\oplus \simeq 0.34 \pm 0.14
\end{equation}
where the uncertainty is from the combination of the scatter in table entries (0.09) and the projection error in the model (0.11).
Thus about one-third of all stars are expected to have a terrestrial-radius planet in the star's HZ.

\section{Discussion}\label{discussion}

\begin{figure}
\centerline{ \plotone{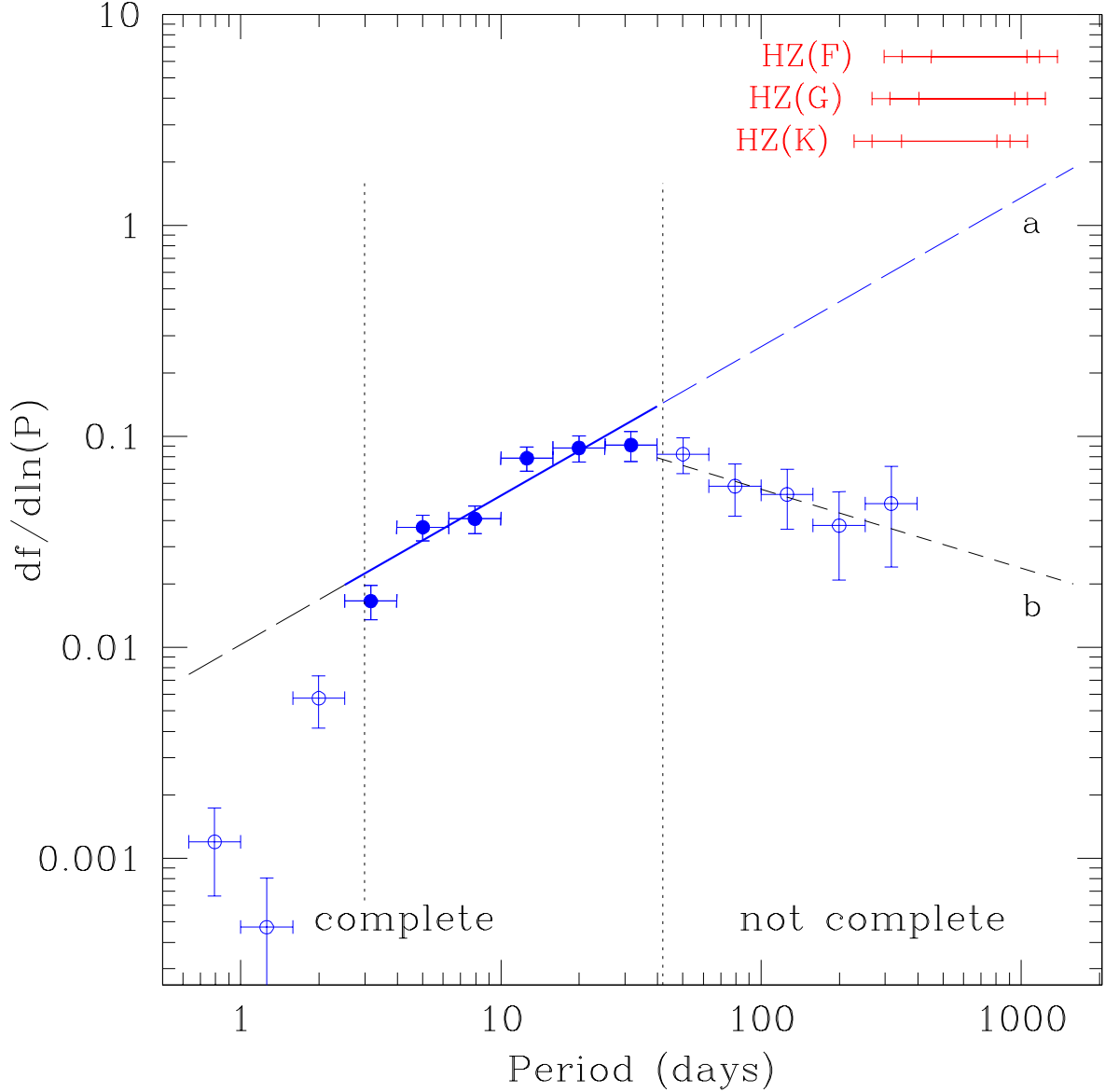} }
 \caption{ The distribution of planets in the population is shown as a function of period.  The distribution is based on a projection from bright stars in the sample database, using the probability of transit as a projection factor for each planet. The data is from Table~\ref{table-p-under}.    In the 3 to 42 day range, the bins are fit by a power law $dN/d\ln{P} \sim P^\beta$ with $\beta = 0.71 \pm 0.08$ (thick line), and extrapolated to longer periods (upper dashed line, labeled ``a'').   The habitable zone ranges for FGK stars are indicated.  The integrated number of planets in these ranges, multiplied by the fraction of terrestrial planets, gives the estimated value of $\eta_\oplus$. The lower dashed line, labeled ``b'', is a fit to the data with periods $>42$ days, however this data is not complete, so the projection is not expected to be a true representation of the distribution in the population. }
 \label{fig-npunder3}
\end{figure}

The projected power law is shown in Fig.~\ref{fig-npunder3} as the dashed line labeled ``a'', based on the best information currently available from Kepler.  On the other hand, if the advice of B2011 is ignored, and the implied populations with periods greater than 42 days are taken seriously, then the dashed line ``b'' would be relevant instead.  Since line ``b'' lies about a factor of 30 below line ``a'', the corresponding value of $\eta_\oplus$ would drop from 34\% to about 1.1\%.

Recently Catanzarite and Shao (2011) estimated $\eta_\oplus \simeq (1 - 3)\%$, using the same B2011 database, but making the fundamentally different assumption that periods greater than 42 days are as valid as shorter ones. Also, they did not compensate for the bias against small planets around faint stars.   Their assumptions stand in marked contrast to those in the present paper.  Their assumptions also disagreed with  the statements in B2011 that the data is not complete beyond 42 days, since those longer periods were looked for in the range beyond the first 136 days in an \textit{ad hoc}, i.e., not a systematic and complete fashion.  The large difference, $(1 - 3)\%$ versus $(34 \pm 14)\%$, illustrates why it would be valuable if the Kepler mission could be extended in time, so as to be able to make measurements in the HZ range of periods, bypassing the current need to extrapolate to these periods.

Another point of comparison might be to ask for the value of $df/d\ln P$ for terrestrial planets in the HZ in the Solar System.  Taking $P_1 = 224$ days for Venus, and $P_2 = 686$ days for Mars, and assuming that the encompassed 3 planets effectively have 2 planets between these limits, and assuming that every star has such a planet system, a frequency value of
\begin{equation}
  \eta_\oplus(SS) \simeq  (df/d\ln P)_{SS} \approx (3-1)/\ln(686/224) \simeq 1.8
\end{equation}
is found.  Thus the Kepler value of about 5 times smaller is not too surprising, especially considering that, for short-period planets, there is only about 1 planet for every 3 or 4 stars (i.e., $1/0.29 \simeq 3.4$) in the population (Sec.~\ref{f-spty}).  This comparison also suggests that the projected line ``a'' in Fig.~\ref{fig-npunder3} is consistent with a density of planets per individual star that does not exceed dynamical limits, given that the inner Solar System is believed to be dynamically stable.

\section{Conclusions}\label{conclusions}

In the current Kepler database (B2011), transits with periods less than 42 days for bright, ``Sun-like'' FGK target stars are analyzed in order to estimate the frequency of terrestrial, habitable-zone planets in the target population, giving $\eta_\oplus \simeq (34 \pm 14)\%$.  The quoted uncertainty is the formal error in projecting the numbers of short-period planets.  The true uncertainty will remain unknown until Kepler observations of orbital periods in the 1000-day range become available.

\acknowledgments


I thank the Kepler Team for providing such abundant and precise data, and for helpful comments on this paper.
I thank the staff at the Computation Facility of the Harvard-Smithsonian Center for Astrophysics.
Finally, I thank the referees, Jim Kasting and anonymous, who made especially useful comments, and who therefore had a key influence on the final version of this paper.
Part of this research was carried out at the Jet Propulsion Laboratory, California Institute of Technology, under a contract with the National Aeronautics and Space Administration.

\clearpage  

\begin{deluxetable}{ccccccccc}  
\tabletypesize{\small}
\tablecaption{Radius \textit{vs} $K_p$ in Sample \label{table-nnkp}}
\tablewidth{0pt}
\tablehead{ $K_p$ range & $r=$  & $r=$    & $r=$    & $r=$    &  $r=$   &  $r=$    & $N_s$ & n(mid-r)/$N_s$ \\
                (mag)   & all-r & 0.6-1.4 & 1.5-2.0 & 2.1-3.0 & 3.1-8.0 & 8.1-39.7 &       &                 }
\startdata
10.0, 12.999 & 136 & 25 & 32  &  33  & 38 &  8  &  21,822  &  $0.47 \pm 0.05\%$  \\
13.0, 13.999 & 213 & 33 & 47  &  64  & 47 & 22  &  34,903  &  $0.45 \pm 0.04\%$  \\
14.0, 14.999 & 315 & 25 & 65  & 110  & 78 & 37  &  39,285  &  $0.64 \pm 0.04\%$  \\
15.0, 15.999 & 288 & 13 & 57  &  89  & 78 & 51  &  55,952  &  $0.40 \pm 0.03\%$  \\
\enddata 

Notes:
For each magnitude interval (col.1), the table lists the number of Kepler transits around FGK stars in the sample,
for all planet radii (col. 2),
and for 5 sub-ranges of radii (cols. 3-7).
The number of basis stars $N_s$ is in col. 8, and the fraction of stars with mid-size
planets detected, in each magnitude bin, is in col. 9.
All periods $P < 42$ days are included.

\end{deluxetable}

\begin{deluxetable}{crrr}  
\tabletypesize{\small}
\tablecaption{Radius Distribution in Sample \label{table-r}}
\tablewidth{0pt}
\tablehead{ $log(r)$  &  $n_p$        &  $n_p$     &  $n_p$    \\
             (range)  &  (all $K_p$)  &  (bright)  &  (faint)   }
\startdata
-0.30, -0.15 &   8  &    6  &    2    \\
-0.15, 0.00  &  13  &    9  &    4    \\
0.00, 0.15  &   76  &   44  &   32    \\
0.15, 0.30  &  161  &   59  &  102    \\
0.30, 0.45  &  302  &  107  &  195    \\
0.45, 0.60  &  146  &   50  &   96    \\
0.60, 0.75  &   74  &   32  &   42    \\
0.75, 0.90  &   57  &   18  &   39    \\
0.90, 1.05  &   53  &   15  &   38    \\
1.05, 1.20  &   47  &    9  &   38    \\
1.20, 1.65  &   22  &    6  &   16    \\
totals      &  959  &  355  &  604    \\
\enddata 

Notes:
For each radius interval (col.1), the table lists the number of Kepler planets around FGK stars in the sample,
for all Kepler magnitudes (col. 2),
and for each of the bright (col. 3) and faint (col. 4) ranges.
The last bin in $r$ is wider than the others.
All periods $P < 42$ days are included.

\end{deluxetable}

\begin{deluxetable}{cccc}  
\tabletypesize{\small}
\tablecaption{Period Distribution in Sample \label{table-p}}
\tablewidth{0pt}
\tablehead{  $\log(P)$  &  $n_p$       &  $n_p$    &  $n_p$    \\
               (range)  & (all $K_p$)  &  (bright) &  (faint)   }
\startdata
-0.50, 0.00 &   11  &   5  &   6    \\
0.00, 0.25  &   23  &   5  &  18    \\
0.25, 0.50  &   61  &  16  &  45    \\
0.50, 0.75  &  161  &  58  & 103    \\
0.75, 1.00  &  192  &  64  & 128    \\
1.00, 1.25  &  182  &  71  & 111    \\
1.25, 1.50  &  154  &  62  &  92    \\
1.50, 1.75  &  100  &  40  &  60    \\
1.75, 2.00  &   32  &  15  &  17    \\
2.00, 2.25  &   29  &  13  &  16    \\
2.25, 2.75  &   13  &   6  &   7    \\
totals      &  958  & 355  & 603    \\
\enddata 

Notes:
Table lists the number of Kepler planets around FGK stars in the sample,
in each period range (col. 1),
for all Kepler magnitudes (col. 2),
for each of the bright (col. 3) and faint (col. 4) ranges,
and for periods $P < 42$ days.
The first and last bins are wider than the others.

\end{deluxetable}

\begin{deluxetable}{crrc}  
\tabletypesize{\small}
\tablecaption{Period Distribution in Population \label{table-p-under}}
\tablewidth{0pt}
\tablehead{  $P$    &  $n_p$ &  $N_p$ &  $\frac{N_p/N_s}{\Delta \ln P}$  \\
          (range)   &           &           &                            }
\startdata
0.63, 1.00   &   5  &    19.8  &  0.00120 \\
1.00, 1.58   &   2  &     7.8  &  0.00047 \\
1.58, 2.51   &  13  &    94.9  &  0.00574 \\
2.51, 3.98   &  29  &   274.1  &  0.01658 \\
3.98, 6.31   &  52  &   613.3  &  0.03710 \\
6.31, 10.0   &  46  &   672.5  &  0.04068 \\
10.0, 15.8   &  59  &  1302.1  &  0.07877 \\
15.8, 25.1   &  51  &  1457.0  &  0.08814 \\
25.1, 39.8   &  38  &  1503.1  &  0.09093 \\
39.8, 63.1   &  27  &  1361.6  &  0.08237 \\
63.1, 100.   &  13  &   959.5  &  0.05804 \\
100., 158.   &  10  &   877.6  &  0.05309 \\
158., 251.   &   5  &   624.3  &  0.03777 \\
251., 398.   &   4  &   793.5  &  0.04800 \\
total        & 354  &          &          \\
\enddata 

Notes:
For bright FGK stars ($K_p < 14$),
in each period range (col. 1),
the table lists the number of planets in the sample $n_p$ (col. 2),
the inferred number of planets in the population $N_p$ (col.3), and
the corresponding number of planets per star per bin width in $\log P$.

\end{deluxetable}

\begin{deluxetable}{cccccccccc}  
\tabletypesize{\small}
\tablecaption{Planet \& Star Numbers in Sample \& Population \label{table-pssp}}
\tablewidth{0pt}
\tablehead{  SpTy & $n_p(terr)$ & $N_p(terr)$ & $n_p(ice)$ & $N_p(ice)$ & $n_p(gas)$ & $N_p(gas)$ &  $n_p$  & $N_p$ &  $N_s$  }
\startdata
F   &  46 & 1017 &   53 & 1697 &  11 &  620 &  110 &  3334  & 11819  \\
G   &  65 & 1317 &   96 & 3586 &  12 &  278 &  173 &  5181  & 14997  \\
K   &  29 &  739 &   36 & 1179 &   7 &  138 &   72 &  2056  &  9080  \\
FGK & 140 & 3073 &  185 & 6462 &  30 & 1036 &  355 & 10571  & 35896  \\
\enddata 

Notes:
Col. 1 is the spectral type of host star,
cols. 2-3 are the number of terrestrial planets in the sample and population,
cols. 4-5 are similar for ice giants,
cols. 6-7 are similar for gas giants,
cols. 8-9 are the numbers of planets in the sample and population, and
col. 10 is the number of stars in the population (i.e., bright Kepler stars with periods less than 42 days).
The bottom row is for the sum of all three spectral types.

\end{deluxetable}

\begin{deluxetable}{ccccc}  
\tabletypesize{\small}
\tablecaption{Planet and Star Types in Population \label{table-f-spty}}
\tablewidth{0pt}
\tablehead{  SpTy & $\frac{N_p(terr)}{N_s}$ & $\frac{N_p(ice)}{N_s}$  & $\frac{N_p(gas)}{N_s}$ & $\frac{N_p(all)}{N_s}$ \\
                  &       (\%)              &        (\%)             &      (\%)              &       (\%)              }
\startdata
F   &  $9 \pm 1$  &  $14 \pm 2$  &  $5 \pm 2$ & $28 \pm 3$ \\
G   &  $9 \pm 1$  &  $24 \pm 2$  &  $2 \pm 1$ & $35 \pm 3$ \\
K   &  $8 \pm 2$  &  $13 \pm 2$  &  $2 \pm 1$ & $23 \pm 3$ \\
FGK &  $9 \pm 1$  &  $18 \pm 1$  &  $3 \pm 1$ & $29 \pm 2$ \\
\enddata 

Notes:
Col. 1 is the spectral type of host star,
cols. 2-4 are the ratios (\%) of planets (terrestrial, ice giant, and gas giant) to stars (F, G, K, and FGK) in the population,
including uncertainties,
and col. 5 is the ratio (\%) of all planets to each star type.
Data are from bright Kepler stars with periods less than 42 days.

\end{deluxetable}

\begin{deluxetable}{cccccc}  
\tabletypesize{\small}
\tablecaption{Habitable-Zone Periods \label{table-HZ}}
\tablewidth{0pt}
\tablehead{ HZ type  & Characteristic  &  $a_\odot$ range  &  P(F) range  &  P(G) range &  P(K) range \\
                     &                 &  (AU)             &    (days)    &   (days)    &   (days)     }
\startdata
Case 1  &  wide     &  0.72-2.00  &  297-1377  &  267-1238  &  228-1057  \\
Case 2  &  nominal  &  0.80-1.80  &  348-1176  &  313-1057  &  267-903   \\
Case 3  &  narrow   &  0.95-1.67  &  451-1050  &  405-944   &  346-807   \\
\enddata 

Notes:
Cols. 1-2 list the case number and one-word description of the three types of HZ in this paper,
col. 3 gives the Sun-planet separation range for each Case, and
cols. 4-6 give the corresponding period ranges for FGK stars.

\end{deluxetable}

\begin{deluxetable}{cccc}  
\tabletypesize{\small}
\tablecaption{Terrestrial HZ Planets in Population \label{table-eta-terr}}
\tablewidth{0pt}
\tablehead{ HZ type  &  $\eta_\oplus(F)$  &  $\eta_\oplus(G)$  & $\eta_\oplus(K)$  }
\startdata
Case 1  &  0.47  &  0.44  &  0.39    \\
Case 2  &  0.37  &  0.34  &  0.31    \\
Case 3  &  0.27  &  0.25  &  0.22    \\
\enddata 

Notes:
Col. 1 is the HZ case number as described in the text.
Cols. 2-4 give the expected number of terrestrial-radius planets, per star, in the HZ, for each spectral type.

\end{deluxetable}

\end{document}